\documentclass[aps,pre,twocolumn,superscriptaddress,showpacs]{revtex4-1}

\usepackage{epsfig}
\usepackage{amsmath,amssymb,bm}
\usepackage{graphics}
\usepackage{epsfig}
\usepackage{graphicx}

\begin{document}

%%\title{Cores structure of directed complex networks}
\title{$\textbf{k}$-core percolation on multiplex networks
%%$e$-
}

\author{N. Azimi-Tafreshi}

\affiliation{Physics Department, Institute for Advanced Studies in Basic Sciences, 45195-1159 Zanjan, Iran }

\author{J. G\'omez-Garde\~nes}
\affiliation{Departamento de F{\'\i}sica de la Materia Condensada, Universidad de Zaragoza, E-50009 Zaragoza, Spain}
\affiliation{Instituto de Biocomputaci\'on y F{\'\i}sica de los Sistemas Complejos (BIFI), Universidad de Zaragoza, E-50018 Zaragoza, Spain}

\author{S.~N. Dorogovtsev}
\affiliation{Departamento de F{\'\i}sica da Universidade de Aveiro $\&$ I3N, Campus Universit\'ario de Santiago, 3810-193 Aveiro, Portugal}
\affiliation{A.F. Ioffe Physico-Technical Institute, 194021 St. Petersburg, Russia}

%%\date{}

%%\maketitle
\begin{abstract}
We generalize the theory of $k$-core percolation on complex networks to $\textbf{k}$-core percolation on multiplex networks, where $\textbf{k}\equiv (k_a, k_b, \ldots)$. Multiplex networks can be defined as networks with a set of vertices but different types of edges, $a,b,\ldots$, representing different types of interactions. For such networks, the $\textbf{k}$-core is defined as the largest subgraph in which each vertex has at least $k_i$ edges of each type, $i=a,b,\ldots$. We derive self-consistency equations to obtain the birth points of the $\textbf{k}$-cores and their relative sizes for uncorrelated multiplex networks with an arbitrary degree distribution. To clarify our general results, we consider in detail multiplex networks with
edges of two types, $a$ and $b$, and solve the equations in the particular case of Erd\H{o}s--R\'enyi and scale-free multiplex networks. %%At the emergence point of $\textbf{k}$-cores, a hybrid phase transition occurs, except for $(1,1)$-core in which the transition is continuous.
We find hybrid phase transitions at the emergence points of $\textbf{k}$-cores except the $(1,1)$-core for which the transition is continuous.
%%for values $k_i\geq 2$.
%%As an application to the real networks,
We apply the $\textbf{k}$-core decomposition algorithm to air-transportation multiplex networks, composed of two layers, and obtain the size of $(k_a, k_b)$-cores.
\end{abstract}


\maketitle

\section{Introduction}
\label{s1}
%\textit{Introduction}.---

In the last decade, during the advent of network science, a number of statistical descriptions were proposed to characterize the structure of the interactions of many diverse complex systems \cite{struc1,struc2,struc3,struc4}. One of the most fundamental features when characterizing the structure of a network is the size of its giant connected component, {\em i.e.}, the size of the largest subset of vertices so that each pair of them is connected by a path of finite length. The existence of a giant connected component is particularly important in those networks carrying flows of different natures, such as viruses and rumours in social systems, data in technological networks or goods and humans in transportation systems. Thus, the maximum capacity for the spread and transport in such systems is limited by the size of the giant connected component. Apart from this practical meaning, the giant connected component is the relevant order parameter that shows the formation of a macroscopic cluster in the context of ordinary percolation \cite{giant,Dorogovtsev:dm-books,Dorogovtsev:dgm08}.

Apart from ordinary percolation, which leads to a continuous phase transition for the emergence of a macroscopic giant connected component, a number of generalizations to the usual scenario were introduced. These variants lead to other kinds of giant connected components, whose emergence is associated with different phase transitions \cite{k-core1,k-core2,k-clique1,k-clique2,explosive1,explosive2}. Among these generalizations is $k$-core percolation, in which a giant $k$-core exists if the vertex mean degree of the network exceeds some threshold \cite{k-core1,k-core2}. The $k$-core of a complex network is defined as the largest subgraph whose vertices have degree at least $k$. Thus, each two vertices in the $k$-core are interconnected by at least $k$ paths, which may partially overlap with each other.

The $k$-core of a given graph can be obtained by a recursive pruning algorithm that, at each step, removes all existing vertices with degrees less than $k$. As the result of this pruning, the network is decomposed into a hierarchical set of progressively enclosed $k$-cores with the highest $k$-core being placed in the center. The application of this graph decomposition technique to large real-world networks allows to describe qualitatively their structure in terms of the complete set of their $k$-cores  \cite{Alvarez:decomposition, k-shell:Jellyfish}. Moreover, it was shown that the
most efficient vertices in spreading processes are those belonging to higher $k$-cores of a network \cite{k-shell:spreaders}. For these reasons, during the last years the $k$-core organization of complex networks has been extensively studied \cite{k-core1,k-core2,Baxter:k-core}. The most remarkable result is that for $k\geq 3$, a $k$-core emerges discontinuously at the percolation threshold through a hybrid phase transition, combining a discontinuity and a critical singularity and thus breaking the usual continuous scenario of ordinary percolation.

Very recently, it has been considered that most of real-world networks are not isolated objects but composed of many coupled, interdependent networks
such that the function of one network depends on the others \cite{Interdependent1, Interdependent2}. For such networks, the pruning of vertices in one network can lead to removal of dependent vertices in other networks. It was found that generalized percolation properties of these interdependent networks differ strongly from ordinary
percolation on a single network \cite{Buldyrev:cascade, inter:percolation1, inter:robustness, inter:percolation2}. In fact, they more resemble features of k-core percolation
in single networks including the hybrid phase transition.

The simplest particular case
%%representation
of interdependent networks are multiplex networks in which each vertex depends on at most one vertex of other networks. Multiplex networks can be treated as
%%a overlapping
superpositions of several
%%layer of
networks (sometimes called layers) with different edges \cite{ML1}. In other words, all vertices in these networks are of the same type, but the edges are of different types (colors). Note that here a vertex not necessary has all types of connections. Multiplexes have recently attracted a lot of attention as they are the kind of substrates representing better the interaction patterns occurring in many real systems, in which several ways for the interaction between the elements coexist. This is the case of transportation, social and technological networks among others.

Based on its ubiquity, the structural and dynamical properties of multiplexes have been studied in several contexts including diffusion \cite{ML2,ML5}, evolutionary games \cite{ML3}, Boolean dynamics \cite{multiplex2}, epidemics \cite{ML4} and, of course,  percolation \cite{multiplex1,Baxter:Avalanche}. In this way, in \cite{Baxter:Avalanche}, a giant viable component for multiplex networks is defined as a set of vertices in which, for every type of edges, each two vertices are interconnected by at least one path following only edges of this type. Clearly, the giant viable component of a multiplex network is a subgraph of the giant connected components of all single networks (layers). The theory of ordinary percolation on multiplex networks shows a hybrid transition which is the birth of the giant viable components, similar to what happens in $k$-core percolation on single networks. Furthermore, in \cite{Buldyrev:cascade} it was shown that the giant viable component is a subgraph of the so-called mutual component (or mutually connected component). By definition, a vertex belongs
to the mutual component if for every color of edges, which he has, at least one its nearest neighbor connected by an edge of this color is in the mutual component.
Since vertices may have among their edges, some of colors missed, some pairs of vertices in the mutual component may be interconnected by paths of incomplete set of the colors.
%, composed of those vertices that in each network layer, in which they share connections of the corresponding type, they are connected to at least one neighbor belonging to this mutual component. Therefore, some pairs of vertices in the mutual component may be interconnected by paths of incomplete set of colors.

In this paper we study the organization of specific subgraphs, $\textbf{k}$-cores for multiplex networks, where $\textbf{k}\equiv (k_a, k_b, \ldots)$. Generalizing the notion of a viable connected component, we define the $\textbf{k}$-viable component as a set of vertices, in which for each type $i$ of edges and for each two vertices, there are at least $k_i$ paths, following only edges of type $i$. The $\textbf{k}$-core is a giant $\textbf{k}$-viable component,
where we must assume for each type $i$ of edges that $k_{i}\geq 2$.

The paper is organized as follows. In section~\ref{s2}, we introduce an algorithm for the $\textbf{k}$-core decomposition of multiplex networks and present an analytical framework enabling us to describe the nature of the transitions corresponding to the emergence of $\textbf{k}$-cores with arbitrary $(k_a, k_b, \ldots)$. We apply our general results to the Erd\H{o}s--R\'enyi and scale-free multiplex networks. In section~\ref{s3}, as an application to the real multiplexes, we apply the $\textbf{k}$-core algorithm to air-transportation multiplexes and compare the results with our analytical predictions.

%%The paper is concluded in section IV.
%%%%%%%%%%%%%%%%%%%%%%%%%%%%%%%%
%%%%%%%%%%%%%%%%%%%%%%%%%%%%%%%%
%%%%%%%%%%%%%%%%%%%%%%%%%%%%%%%%
\section{k-core on multiplex networks}
\label{s2}

%%%%%%%%%%%%%%%%
%%%%%%%%%%%%%%%%

\subsection{Analytical framework}

\begin{figure}[t!]
\begin{center}
\scalebox{0.32}{\includegraphics[angle=0]{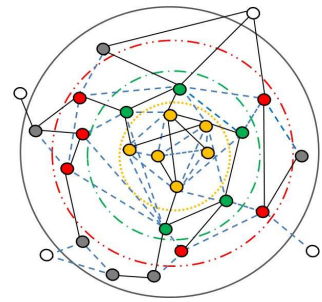}}
\end{center}
\caption{The $(k_a,k_b)$-core decomposition for a multiplex network with two types of edges. The cores from out to center are the $(1,1)$-core, the $(1,2)$-core, the $(2,2)$-core, and $(1,3)$-core.
}
\label{f1}
\end{figure}

%\textit{Definition}.---
Let us consider a multiplex network, having edges of types $a,b, \ldots$, with a given joint degree distribution $P(q_a,q_b,\ldots)$ and a locally tree-like structure in the infinite network limit. For simplicity, we assume that this network is completely uncorrelated, though, in principle, correlations between different edges, $i=a,b,\ldots$, of a vertex might be easily taken into account.
%%connections of different type of
%%each of layers are uncorrelated networks, but in general different connections of each vertex can be correlated. The $\textbf{k}$-core of this network, is defined as the largest subgraph in which each vertex has at least $k_i$ edges
%%of all types $i=a, b, \ldots$.
To obtain the $\textbf{k}$-core of a multiplex network, we use the following pruning algorithm: at each step we remove every vertex if for at least one type of edge $i$, $q_i<k_i$. As the result of the pruning, the degrees of some vertices change. If there are still vertices which one can prune, we remove them in the next step. The pruning is continued until no vertex remains of degree $q_i$ less than the threshold $k_i$. Fig.~\ref{f1}  shows the $\textbf{k}$-core decomposition for a multiplex network with two types of edges.
%%%%%%%%%%%%%%%%%%%%%%%%%%%%%%%%%%%%%%%%%%%%%%%%%%%%%%%%%
%%%%%%%%%%%%%%%%%%%%%%%%%%%%%%%%%%%%%%%%%%%%%%%%%%%%%%%%%
%%%%%%%%%%%%%%%%%%%%%%%%%%%%%%%%%%%%%%%%%%%%%%%%%%%%%%%%%
%%%%%%%%%%%%%%%%%%%%%%%%%%%%%%%%%%%%%%%%%%%%%%%%%%%%%%%%%
%\textit{Basic Equations}.---

\begin{figure}[t!]
\begin{center}
\scalebox{0.35}{\includegraphics[angle=0]{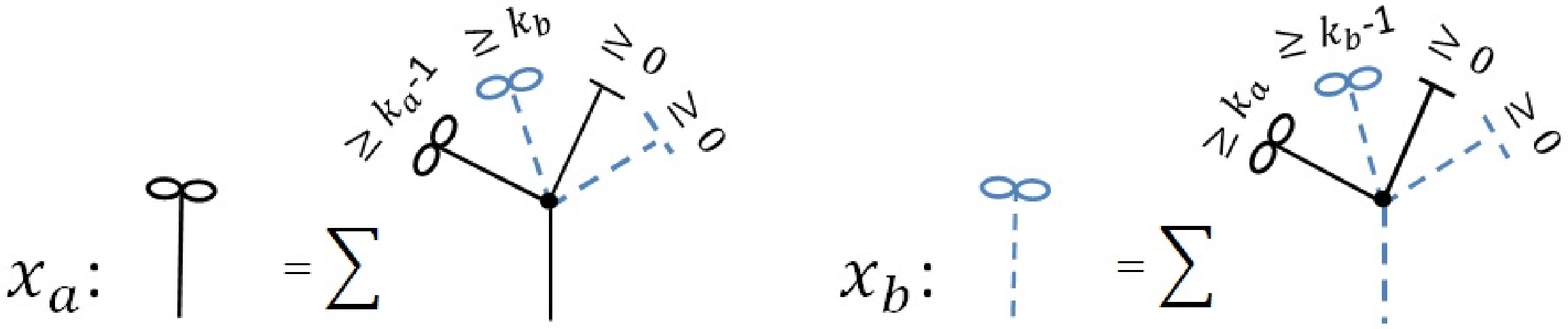}}
\end{center}
\caption{Schematic representation of the self-consistency equations for the probabilities $x_a$ and $x_b$. The solid black and dashed blue lines with infinity symbols at one of their ends, represent probabilities $x_a$ and $x_b$ respectively. The edges lead to finite components, namely $1-x_a$ and $1-x_b$, are shown by solid and dashed lines with cuts at one of their ends.}
\label{f00}
\end{figure}

To find the size of the giant $\textbf{k}$-cores, for each type $i$ of edges, we define $x_i$ as the probability that an end vertex of a randomly chosen edge of type $i$ is the root of an infinite subtree of type $i$. The subtree of type $i$ is, by definition, a tree whose vertices have at least $k_i-1$ edges of type $i$ and at least $k_j$ edges of each of other types $j\neq i$ edges.
%%, which $j\neq i$.
Probabilities $x_a$ and $x_b$ for a multiplex network with two types $a$ and $b$ of edges are schematically shown in Fig.~\ref{f00}.  These probabilities play the role of the order parameters of the phase transition associated with the emergence of the $\textbf{k}$-cores. We can write the self-consistency equations for probabilities $x_i$ using the locally tree structure of the networks,
\begin{eqnarray}
x_i&=&\sum_{\textbf{q}}\frac{q_{i}P(\textbf{q})}{\langle q_{i}\rangle}
\nonumber
\Big[\sum_{s=k_i-1}^{q_i-1}
\left(
                      \begin{array}{c}
                        q_i-1 \\
                        s \\
                      \end{array}
                    \right)
x_{i}^{s}(1-x_i)^{q_i-1-s}\Big]
\nonumber
\\[5pt]
&&\times\prod_{j\neq i}\Big[\sum_{s'=k_j}^{q_j}
\left(
                      \begin{array}{c}
                        q_j \\
                        s' \\
                      \end{array}
                    \right)
x_{j}^{s'}(1-x_j)^{q_j-s'}\Big],
\label{eq1}
\!\!\!\!\!
\end{eqnarray}
%%%%
where $\textbf{q}\equiv(q_a, q_b,\ldots)$ is the degree of a vertex.

Let us briefly explain Eq.~(\ref{eq1}). The probability that the end vertex of a randomly chosen edge of type $i$ has degree $\textbf{q}$, is $q_{i}P(\textbf{q})/\langle q_{i}\rangle$.  The combinatorial multiplier
$\left(
                      \begin{array}{c}
                        m \\
                        n \\
                      \end{array}
                    \right)$
gives the number of ways one can choose $n$ edges from a sample of $m$ edges. At least $k_i-1$ edges of $q_i-1$ edges of type $i$ (other edges than the starting
%%edge)
one) must lead to an infinite subtree of type $i$ (probability $x_i$) and at least $k_j$ edges of each of $q_j$ edges $(j\neq i)$ must lead to the infinite subtrees of type $j$ (probability $x_j$).

%%%%%%%%%%%%%%%%%%%%%%%%%%%%%%%%%%%%%%%%%%%%%%%%%%%%%%%%%%%
%%%%%%%%%%%%%%%%%%%%%%%%%%%%%%%%%%%%%%%%%%%%%%%%%%%%%%%%%%%
%%%%%%%%%%%%%%%%%%%%%%%%%%%%%%%%%%%%%%%%%%%%%%%%%%%%%%%%%%%
%%%%%%%%%%%%%%%%%%%%%%%%%%%%%%%%%%%%%%%%%%%%%%%%%%%%%%%%%%%
%%
%%%%%%%%%%%%%%%%%%%%%%%%%%%%%%%%%%%%%%%%%%%%%%%%%%%%%%%%%%%
%%%%%%%%%%%%%%%%%%%%%%%%%%%%%%%%%%%%%%%%%%%%%%%%%%%%%%%%%%%
\begin{figure}[b!]
%%\begin{center}
\begin{center}
\scalebox{0.4}{\includegraphics[angle=0]{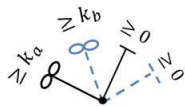}}
\end{center}
\caption{Schematic representation of a vertex belongs to $(k_a,k_b)$-core.}
\label{f0}
\end{figure}
%%%%%%%%%%%%%%%%%%%%%%%%%%%%%%%%%%%%%%%%%%%%%%%%%%%%%%%%%%%
%%%%%%%%%%%%%%%%%%%%%%%%%%%%%%%%%%%%%%%%%%%%%%%%%%%%%%%%%%%

Using these probabilities, we can obtain the relative size of the giant $\textbf{k}$-core. A vertex is in the $\textbf{k}$-core when, for each type of edges $i$, the vertex has at least $k_i$ edges leading to infinite $i$ subtrees. The probability that a vertex belongs to the $(k_a,k_b)$-core for multiplex networks with edges of two types, $a$ and $b$, is shown schematically in Fig.~\ref{f0}. Hence, the relative size of the core is given by the following expression:
\begin{eqnarray}
\!\!\!\!\!\!\!\!\!
n_{\textbf{k}\text{-core}}&=&\sum_{\textbf{q}}P(\textbf{q})
%%\nonumber
\prod_{i}\Big[\sum_{s=k_i}^{q_i}
\left(
                      \begin{array}{c}
                        q_i \\
                        s \\
                      \end{array}
                    \right)
x_{i}^{s}(1-x_i)^{q_i-s}\Big].
%%\nonumber\\
\label{eq2}
\end{eqnarray}
%%%%%%

We can now rewrite Eqs.~(\ref{eq1})--(\ref{eq2}) using generating functions \cite{giant}, which enable us to solve these equations analytically.
%%more simply.
For a single network with a given degree distribution $P(q)$, the generating function $G(x)$ is defined as
\begin{equation}
G
%%_{0}
(x)\equiv \sum_{q} P(q)x^{q}
\label{eq3}
.
\end{equation}
If we assume that there is no correlation between the degrees $q_a, q_b,\ldots$, so that $P(\textbf{q})=P(q_a)P(q_b)\ldots$, then Eqs.~(\ref{eq1}) and (\ref{eq2}) can be rewritten as following
\begin{eqnarray}
x_i&=&\Big[1-\frac{1}{\langle q_i\rangle}\sum_{s=0}^{k_i-2}x_i^s \frac{G_i^{(s+1)}(1-x_i)}{s!}\Big]
\nonumber
\\[5pt]
&&\times\prod_{j\neq i}\Big[1-\sum_{s'=0}^{k_j-1}x_j^{s'} \frac{G_j^{(s')}(1-x_j)}{s'!}\Big]
\label{eq4}
\end{eqnarray}
%%%%%
and
%%%%
\begin{eqnarray}
n_{\textbf{k}\text{-core}}=\prod_i\Big[1-\sum_{s=0}^{k_i-1}x_i^s \frac{G_i^{(s)}(1-x_i)}{s!}\Big]
,
\label{eq5}
\end{eqnarray}
%%%%%%
where we used the notation $G^{s}(x)$ for the $s$-th derivatives of $G(x)$, which produce the higher moments of the distribution $P(q)$.

In general, Eq.~(\ref{eq4}) is a set of self-consistency equations of the following form $x_i= f_i(x_a, x_b, \ldots)$. A hybrid transition occurs at the point in which $f_i$ first meets $x_i$. This point is determined by the condition $\det[\textbf{J}-\textbf{I}]=0$ for the Jacobian matrix $\textbf{J}$, defined as $J_{ij}=\partial f_j/\partial x_i$, and $\textbf{I}$ is the identity matrix. Expanding $f_i$ around the transition point, we find the
singularity of $x_i$ and hence $n_{\textbf{k}\text{-core}}$, which is the relative size of the
%%giant
$\textbf{k}$-core. If we choose $\langle q_i\rangle$ as control parameter, then the singularity at the critical point is
\begin{eqnarray}
\!\!\!\!\!\!\!\!\!
n_{\textbf{k}\text{-core}}-n_{\textbf{k}\text{-core}}^{c}\propto (x_i -x_i^{c})\propto (\langle q_i\rangle -\langle q_i\rangle^{c})^{1/2}
.
\label{eq11}
\end{eqnarray}

The structure of the $\textbf{k}$-core is described by the degree distribution $P_{\textbf{k}}(\textbf{q})$ defined as the probability to find a vertex of degree $\textbf{q}\equiv(q_a,q_b,\ldots)$ in the $\textbf{k}$-core:
%%%%%%%%%%%%
\begin{equation}
P_{\textbf{k}}(\textbf{q})=\frac{n_{\textbf{k}}(\textbf{q})}{n_{\textbf{k}\text{-core}}}
,
\label{eq1new}
\end{equation}
%%%%%%%%%%%%
%%%%%%%%%%%
where $n_{\textbf{k}}(\textbf{q})$ is the fraction of vertices with degree $\textbf{q}$, which fall into the $\textbf{k}$-core. This fraction is given by the following expression:
%%and obtained by the equation:
%%%%%%%%%%%%
\begin{eqnarray}
n_{\textbf{k}}(\textbf{q})&=&\sum_{\textbf{q}' \geq \textbf{q}}P(\textbf{q}')
\nonumber
\prod_{i}\Big[\left(
                      \begin{array}{c}
                        q'_i \\
                        q_{i} \\
                      \end{array}
                    \right)
x_{i}^{q_{i}}(1-x_i)^{q'_i-q_{i}}\Big].\nonumber\\
\label{eq2new}
\end{eqnarray}
%%%%%%%%%%%%
In particular, Eq.~(\ref{eq2new}) helps us to find the size of the corona (a subset of vertices of degree $\textbf{k}$ in the $\textbf{k}$-core, which generalizes the notion of the corona clusters of the $k$-core).
%%Hence, for
Setting $\textbf{q}=\textbf{k}$ and making use of
%%the
generating functions, we find the relative size of the corona:
\begin{eqnarray}
n_{\textbf{k}}(\textbf{k})=\prod_i x_i^{k_{i}} \frac{G_i^{(k_{i})}(1-x_i)}{k_{i}!}.
\label{eq5new}
\!\!\!\!\!
\end{eqnarray}
Furthermore, the
%%total mean vertex
average total degree $q_a+q_b+\ldots$ of a vertex in the $\textbf{k}$-core can be obtained using the degree distribution of the $\textbf{k}$-core in the following way:
%\begin{eqnarray}
%\langle q_i \rangle= \sum_{\textbf{q}}q_iP_{\textbf{k}}(\textbf{q}).
%\label{eq3new}
%\!\!\!\!\!
%\end{eqnarray}
%%%%%
%%%%
\begin{eqnarray}
c_{\textbf{k}\text{-core}}= \sum_{\textbf{q}}(q_a+q_b+\ldots )P_{\textbf{k}}(\textbf{q}).
\label{eq3new}
\end{eqnarray}
%%%%%%%%
To clarify our results, in the following we will consider Erd\H{o}s--R\'enyi and scale-free multiplex networks with two types, $a$ and $b$, of edges to describe the $\textbf{k}$-core organization of these multiplex networks.
%\begin{figure}[t]
%\begin{center}
%\scalebox{0.38}{\includegraphics[angle=0]{kakbfxplot.eps}}
%\end{center}
%\caption{Graphical representation of Eq.~(\ref{eq6}). The largest root of this equation provides the solution of the problem. The non-zero solution exists when $c$ exceeds the critical value $3.8166\ldots$.
%}
%\label{f1}
%\end{figure}
%%%%%%
%%%%%%

%%%%%%%%%%%%%%%%%%%%%%
%%%%%%%%%%%%%%%%%%%%%%

\subsection{Erd\H{o}s--R\'enyi networks}

%\textit{Erd\H{o}s--R\'enyi} (ER) networks.---
Let us first consider Erd\H{o}s--R\'enyi multiplex networks with Poisson degree distributions: $P(q_a)=c_a^{q_a}e^{-c_a q_{a}}/q_{a}!$, $P(q_b)=c_b^{q_b}e^{-c_b q_b}/q_{b}!$, where $c_a$ and $c_b$ are the mean vertex degrees for types $a$ and $b$ of edges, respectively. For the Poisson distribution, the generating function and its $s$-th derivative are $G_i(x)= e^{-c_i (1-x)}$ and $G_i^s(x)= c_i^{s}e^{-c_i(1-x)}$, respectively.

\begin{figure}[t]
\begin{center}
\scalebox{0.40}{\includegraphics[angle=0]{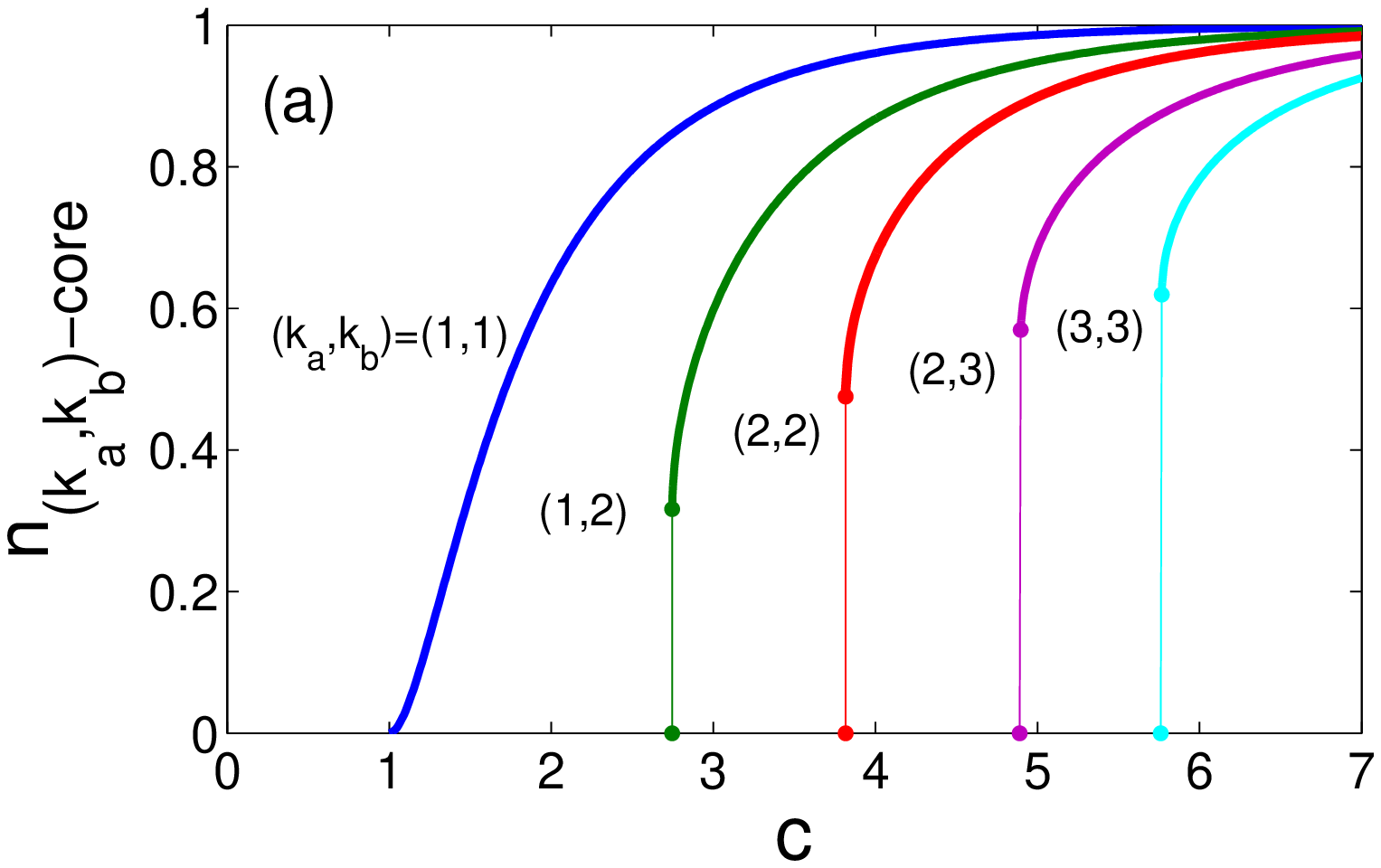}}
\scalebox{0.40}{\includegraphics[angle=0]{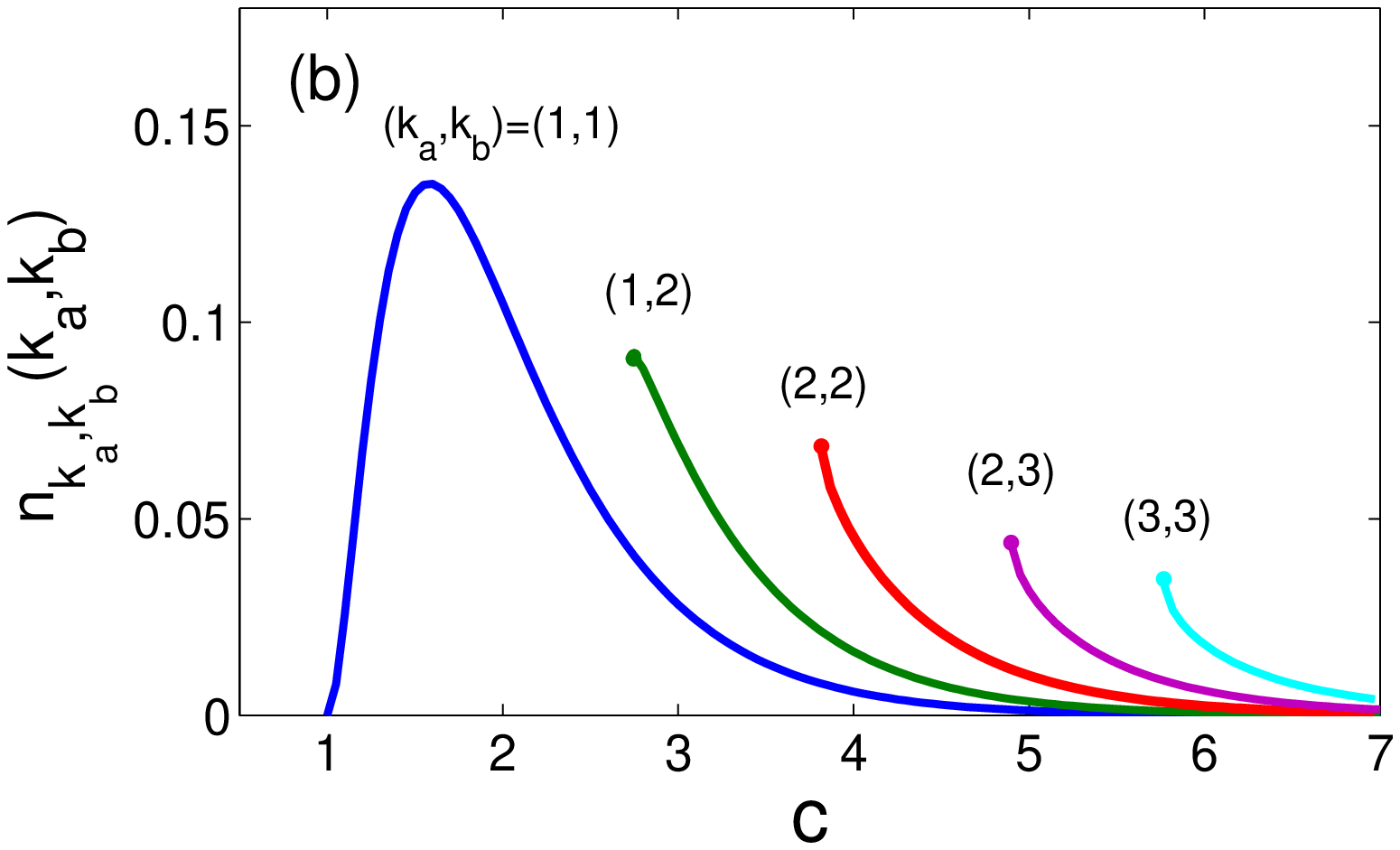}}
\scalebox{0.40}{\includegraphics[angle=0]{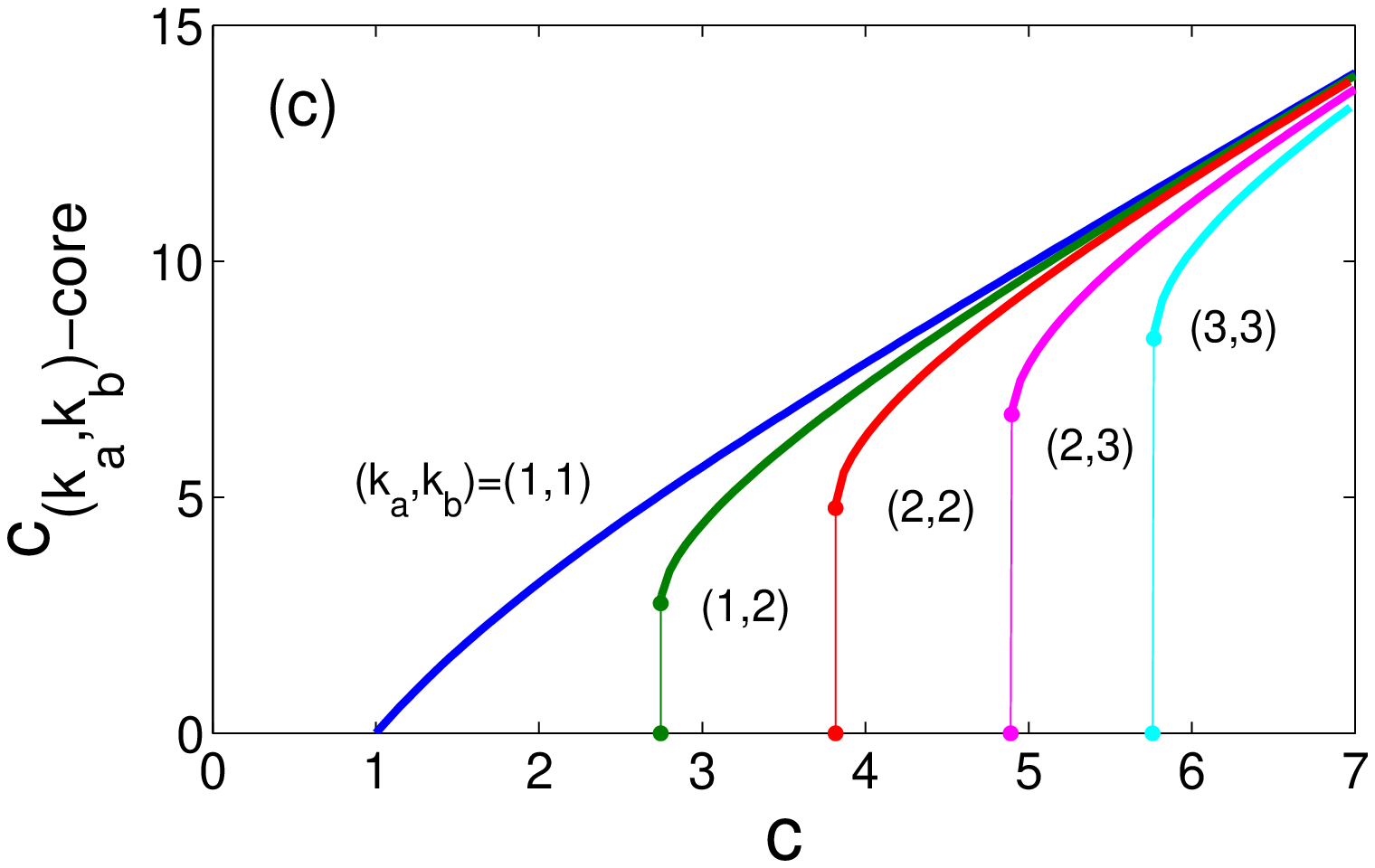}}
\end{center}
\caption{The relative sizes of $(a)$ $\textbf{k}-$core, $(b)$ corona and $(c)$ mean vertex degree of $\textbf{k}-$core in Erd\H{o}s--R\'enyi multiplex networks for some values of $k_a$ and $k_b$, vs the vertex mean degree $c_a=c_b=c$ of the network.
}
\label{f2}
\end{figure}

For the sake of simplicity, let us consider the symmetric case $c_a=c_b\equiv c$. The largest core is the core with $k_a=k_b=1$, that is $(1,1)$-core. In this case
one can obtain $x_a=x_b\equiv X$, such that $X=1-e^{-cX}$.
For $c>1$, this equation has a nonzero nontrivial
%%Increasing the value of $c$, Eq.~(\ref{eq6}) continuously finds a nonzero
solution.
%%at $c=1$.
Figure~\ref{f2}(a) shows the relative size of the
%%giant
$(1,1)$-core, displaying a continuous transition at $c=1$ and is given by the following expression
\begin{eqnarray}
n_{(1,1)\text{-core}}=(1-e^{-cX})^2 \cong 4(c-1)^2,
\label{eq6}
\end{eqnarray}
Notice that the case of the $(1,1)$-core is special since it does not coincide with the viable component
%%, emerged in the ordinary percolation on multiplex networks
\cite{inter:percolation2,Baxter:Avalanche}. Between each two vertices in the $(1,1)$-core, there are at least $2$ distinct paths within the core, following edges that can alternatingly change from one type to the another. Recall that, in contrast to this, between each two vertices in the viable component, for each type of edges, there is at least one path following only edges of that type. The fact that the $(1,1)$-core does not coincide with the viable component is not surprising. Indeed, in ordinary single networks, the $1$-core also does not coincide with the giant connected component ({\em i.e.} the percolation cluster). Instead, it is rather the 2-core that is close to the giant connected component.

%%%%%%%%%%%%%%%%%%%%%%%%%%%%%%%%%%%%%%%%%%%%%%%%%%%%%%%%%%%
%%%%%%%%%%%%%%%%%%%%%%%%%%%%%%%%%%%%%%%%%%%%%%%%%%%%%%%%%%%
%%%%%%%%%%%%%%%%%%%%%%%%%%%%%%%%%%%%%%%%%%%%%%%%%%%%%%%%%%%
%%%%%%%%%%%%%%%%%%%%%%%%%%%%%%%%%%%%%%%%%%%%%%%%%%%%%%%%%%%

Furthermore, if only one of the two $k_i$ values is equal to $1$, the
%%giant
$(k_a,k_b)$-core is still not a $(k_a,k_b)$-viable component in the sense that there are still not $k_a+k_b$ paths between two vertices following edges of the corresponding types. However, in this case there are $k_a+k_b \geq 3$ paths between each two vertices within the $(k_a,k_b)$-core following edges of alternating types. For instance, let us consider the
%%next core that is
$(1,2)$-core. In this case, one can find the following equations for $x_a$ and $x_b$:
\begin{eqnarray}
x_a&=&(1-e^{-c_bx_b}-c_bx_be^{-c_bx_b})
,\nonumber
\label{eq7a}
\\[5pt]
x_b&=&(1-e^{-c_bx_b})(1-e^{-c_ax_a})
.
\label{eq7b}
\end{eqnarray}
Using these probabilities, the relative size of the $(1,2)$-core is given by the following expression:
\begin{eqnarray}
\!\!\!\!\!\!\!\!\!\!
n_{(1,2)\text{-core}}&=&(1-e^{-c_a x_a})
%%\times \nonumber\\&&
(1-e^{-c_b x_b}-c_b x_b e^{-c_b x_b}).
\label{eq7}
\end{eqnarray}
Fig.~\ref{f2}(a), in particular, shows the relative size of the $n_{(1,2)\text{-core}}$ in the symmetric case of $c_a=c_b=c$. While the giant core is non-viable, a hybrid transition appears at the emergence point of the core at $c\simeq 2.7461$.
%%%%%%%%%%%%%%%%%%%%%%%%%%%
%%%%%%%%%%%%%%%%%%%%%%%%%%%%

Let us now assume that each of $k_i$ exceeds $1$.
%%The other case is if non of $k_i$ values is equal to one.
In this case, considering  the tree for $x_i$ in Fig.~\ref{f00}, one can show that this $\textbf{k}$-core is a $\textbf{k}$-viable component in the sense that for every type $i$ of edges, each two vertices within the core are interconnected by at least $k_i$ paths following only edges of this type (see Fig.~\ref{f0}). As an example, let us consider $k_a=k_b=2$. In the symmetric case of $c_a=c_b\equiv c$, we find $x\equiv x_a=x_b$ satisfying the following equation:
\begin{eqnarray}
x=(1-e^{-cx})(1-e^{-cx}-cx e^{-cx})
.
\label{eq77}
\end{eqnarray}
%%%%%
Solving the equation for $x$, we can find that the transition occurs at $c\simeq 3.8166 $.  The relative size of $(2,2)$-core is given by the expression:
%%%%%%%%%%%%
%%%%%%%%%%%%%%
\begin{eqnarray}
n_{(2,2)\text{-core}}=(1-e^{-c x}-cx e^{-c x})^2
.
\label{eq777}
\end{eqnarray}
%%%%%%%%%%%%%%%%%
%%Which collapses
This core emerges discontinuously at the transition point, as shown in Fig.~\ref{f2}(a).
%%%%
%%%%%%%%%%%%%%%%%%%%%%%%%%%%%%%%%%%%%%%%%%%%%%%%%%%%%%%%%%%%%%%
\begin{figure}[t]
\begin{center}
\scalebox{0.40}{\includegraphics[angle=0]{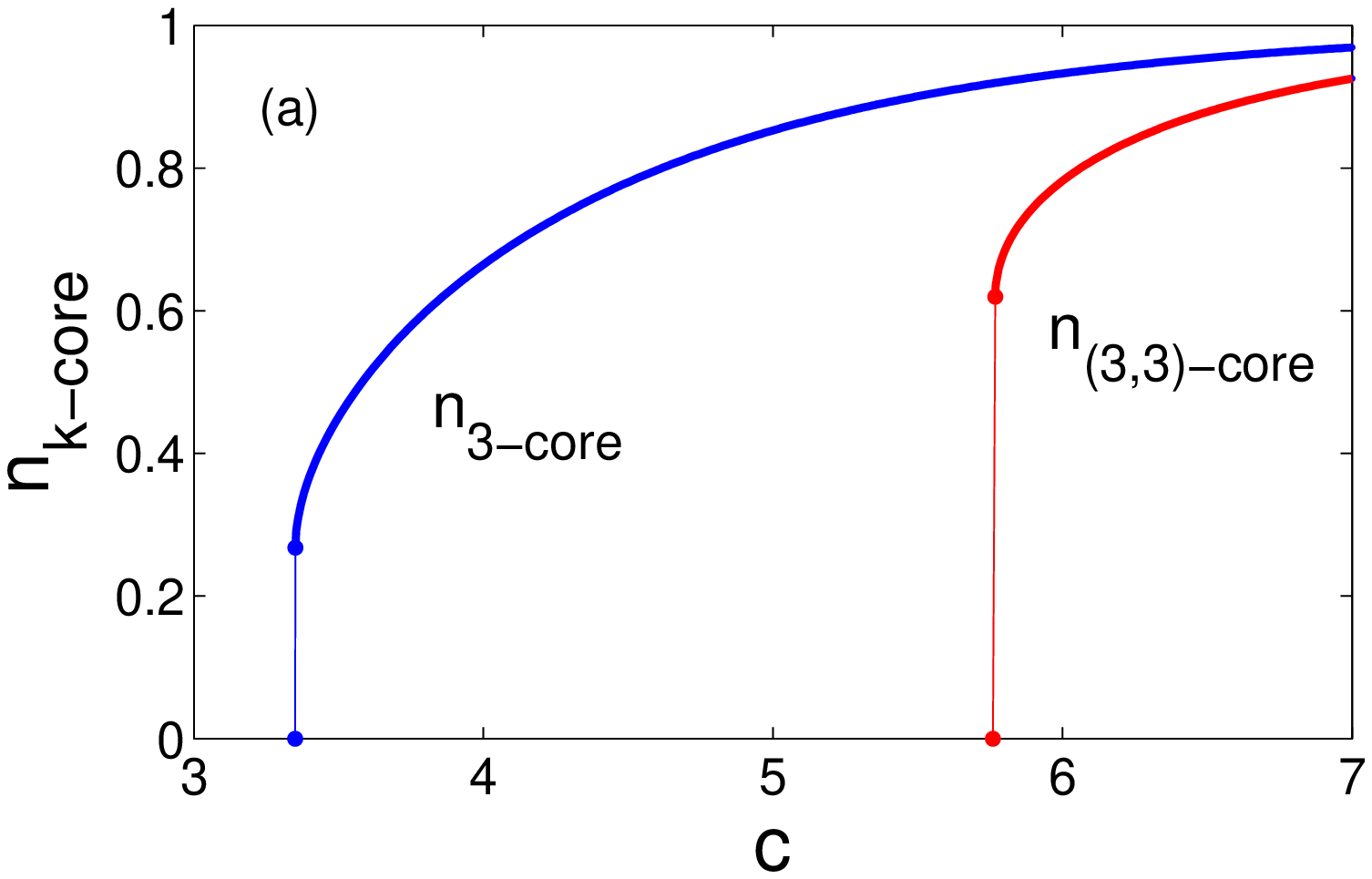}}
\scalebox{0.40}{\includegraphics[angle=0]{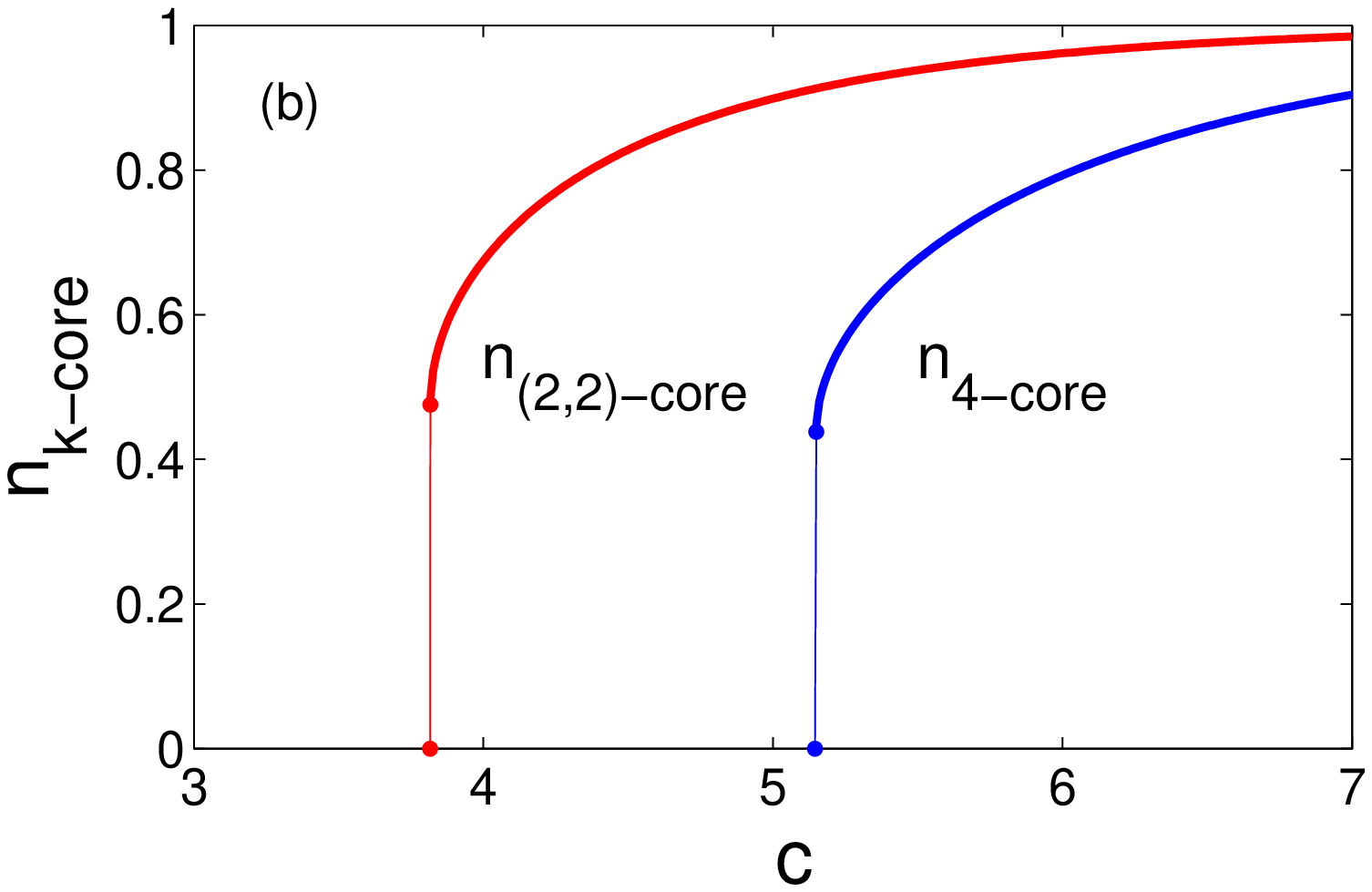}}
\end{center}
\caption{(a) The relative sizes and the birth points of the $(3,3)$-cores in the Erd\H{o}s--R\'enyi multiplex network are compared with the $3$-core in the corresponding ordinary single Erd\H{o}s--R\'enyi networks, (b) the relative size of the $(2,2)$-core of the Erd\H{o}s--R\'enyi multiplex network vs the mean degree of its vertices, compared with the $4$-core in the corresponding ordinary single Erd\H{o}s--R\'enyi networks.}
\label{f33}
\end{figure}
%%%%%%%%%%%%%%%%%%%%%%%%%%%%%%%%%%%%%%%%%%%%%%%%%%%%%%%%%%%%%%%
%%%%
Fig.~\ref{f2}(a) also shows the relative sizes of  the $(k_a,k_b)$-cores for some other values of $k_a$ and $k_b$. There is a jump at the birth points of the cores, which, together with a critical singularity, points out a hybrid transition for $k_a\geq 2$ and $k_b\geq 2$.

One can obtain the relative size of the corona for the Erd\H{o}s--R\'enyi networks using Eq.~(\ref{eq5new}) as
%%%%%%%%%%
\begin{eqnarray}
n_{k_a,k_b}(k_a,k_b)=\frac{(c_{a}x_a)^{k_a}e^{-c_{a}x_a}}{k_{a}!}
%%\times
\, \frac{(c_{b}x_b)^{k_b}e^{-c_{b}x_b}}{k_{b}!}.~~~~~
\label{eq55}
\end{eqnarray}
%%%%%%%%%
Fig.~\ref{f2}(b) displays the dependence of the corona sizes on the vertex mean degree $c$.
Furthermore, using Eq.~(\ref{eq1new}), we can write the degree distribution of the $(k_a,k_b)$-core, that is
%%%%%%%%%%
\begin{eqnarray}
P_{(k_a,k_b)}(q_a,q_b)&=&\frac{1}{n_{(k_a,k_b)\text{-core}}}
\nonumber
\\[5pt]
&\times&\frac{(x_a^{q_a}c_a^{q_a}e^{-c_ax_a})(x_b^{q_b}c_b^{q_b}e^{-c_bx_b})}{q_a!q_b!}
.~~~~
\label{eq9new}
\end{eqnarray}
%%%%%%%%%
%%%%%%%%
Hence, the total vertex mean degree of the $(k_a,k_b)$-core, for uncorrelated Erd\H{o}s--R\'enyi networks with two types of edges $a$ and $b$, is
%%obtained as:
%%
\begin{eqnarray}
c_{(k_a,k_b)}=
%%(
c_a~x_a
%%)
+
%%(
c_b~x_b
%%)
.
\end{eqnarray}
%%%%%%%%%%%
As one can notice from Fig.~\ref{f2}(c), in the symmetric case of $c_a=c_b=c$, the total mean degree of the $(k_a,k_b)$-core changes almost linearly with $c$.

To round off the study of Erd\H{o}s--R\'enyi multiplexes, we compare $\textbf{k}$-core percolation on a Erd\H{o}s--R\'enyi multiplex network and $k$-core percolation on its counterpart single network. As one can see in Fig.~\ref{f33}(a), the $(3,3)$-core emerges at a much higher mean degree value than the $3$-core of the corresponding single Erd\H{o}s--R\'enyi graph. In general, $(k_a,k_b)$-core percolation on the multiplex network has a higher threshold than the $k_a$ or $k_b$-core on single networks. However, the $(k_a+k_b)$-core in a single network has a higher threshold than the $(k_a,k_b)$-core in the corresponding multiplex networks. Fig.~\ref{f33}(b) compares the relative sizes of the $(2,2)$-core of the Erd\H{o}s--R\'enyi multiplex network with the $4$-core in the corresponding ordinary single Erd\H{o}s--R\'enyi networks.
%\begin{figure}[t]
%\begin{center}
%\scalebox{0.40}{\includegraphics[angle=0]{kakb2static.eps}}
%\end{center}
%\caption{Graphical representation of Eq.~(\ref{twodirX}), namely $f(X) \equiv (1-e^{-c X})^2-X=0$. The largest root of this equation provides the solution of the problem. The non-zero solution exists when $c$ exceeds the critical value $0.24554\ldots$.
%}
%\label{f4}
%\end{figure}
%%%%%%%%%%%%%%
%%%%%%%%%%%%%%%%
%\begin{figure}[t]
%\begin{center}
%\scalebox{0.40}{\includegraphics[angle=0]{kakb23static.eps}}
%\end{center}
%\caption{Graphical representation of Eq.~(\ref{twodirX}), namely $f(X) \equiv (1-e^{-c X})^2-X=0$. The largest root of this equation provides the solution of the problem. The non-zero solution exists when $c$ exceeds the critical value $0.24554\ldots$.
%}
%\label{f5}
%\end{figure}

%%%%%%%%%%%%%%%%%%%%%
%%%%%%%%%%%%%%%%%%%%%

\subsection{Scale--free
%%(SF)
networks}
%\textit{Scale--Free (SF) networks.}---

\begin{figure}[t]
\begin{center}
\scalebox{0.45}{\includegraphics[angle=0]{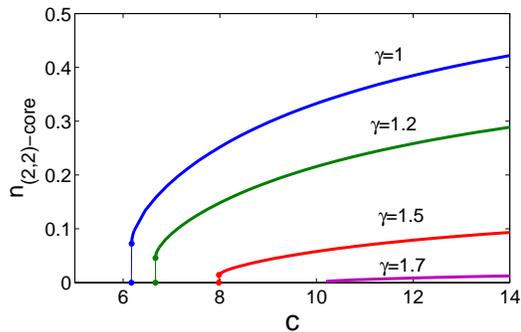}}
\end{center}
\caption{The relative size of the $(2,2)$-core (largest viable $k-$core) for
%%power law degree distributed
scale-free multiplex networks with an exponential degree cutoff at different values of the exponent $\gamma$ of the degree distribution. The core disappears as
%%the value of
$\gamma$ increases and approaches $2$.}
\label{f3}
\end{figure}

%%Now we
Let us consider scale-free multiplex networks with two types of edges. For simplicity we assume that the degrees $q_a$ and $q_b$ are distributed in the same manner with the power-law exponents $\gamma_a=\gamma_b\equiv \gamma$ and the mean degrees $\langle q_a\rangle=\langle q_b\rangle\equiv c$. First we consider the organization of $\textbf{k}$-cores for the power-law distributed networks with an exponential degree cutoff, {\em i.e.}, for instance, $P(q)=\frac{q^{-\gamma}e^{-q/\kappa}}{Li_{\gamma}(e^{-1/\kappa})}$, where $Li_{n}(x)$ is the $n$th polylogarithm of $x$.
%%This distribution is typical for the static model \cite{Goh:static} and it is convenient for analytical consideration.
For this distribution, the generating function defined by Eq.~(\ref{eq3}) is $G(x)=\frac{xLi_{\gamma}(e^{-1/\kappa})}{Li_{\gamma}(e^{-1/\kappa})}$.
The results significantly depend on the low-degree part of the degree distribution.
The presence of a cutoff enables us to consider even low values of $\gamma$, including $\gamma<2$.
For each value of the exponent $\gamma$, we vary the cutoff $\kappa$ (and thus the mean degree $c$) as the control parameter. The relative sizes of the $\textbf{k}$-cores are obtained by solving Eqs.~(\ref{eq4})--(\ref{eq5}). Fig.~\ref{f3} shows $n_{(2,2)\text{-core}}$ for different values of $\gamma$. It becomes clear that the size of the core decreases as $\gamma$ approaches to $2$. Hence, for
%%$SF$
scale-free networks with exponential degree cutoff and, particularly, for the case of pure scale-free networks ($\kappa\rightarrow\infty$), the $(k_a,k_b)$-core does not exist for $\gamma >2$.

%%%%%%%%%%%%%%%%%%%%%%%%%%%%%%%%%%%%%%%%
%%%%%%%%%%%%%%%%%%%%%%%%%%%%%%%%%%%%%%%%
%%%%%%%%%%%%%%%%%%%%%%%%%%%%%%%%%%%%%%%%%
%%%%%%%%%%%%%%%%%%%%%%%%%%%%%%%%%%%%%%%%
\begin{figure}[t!]
\begin{center}
\scalebox{0.45}{\includegraphics[angle=0]{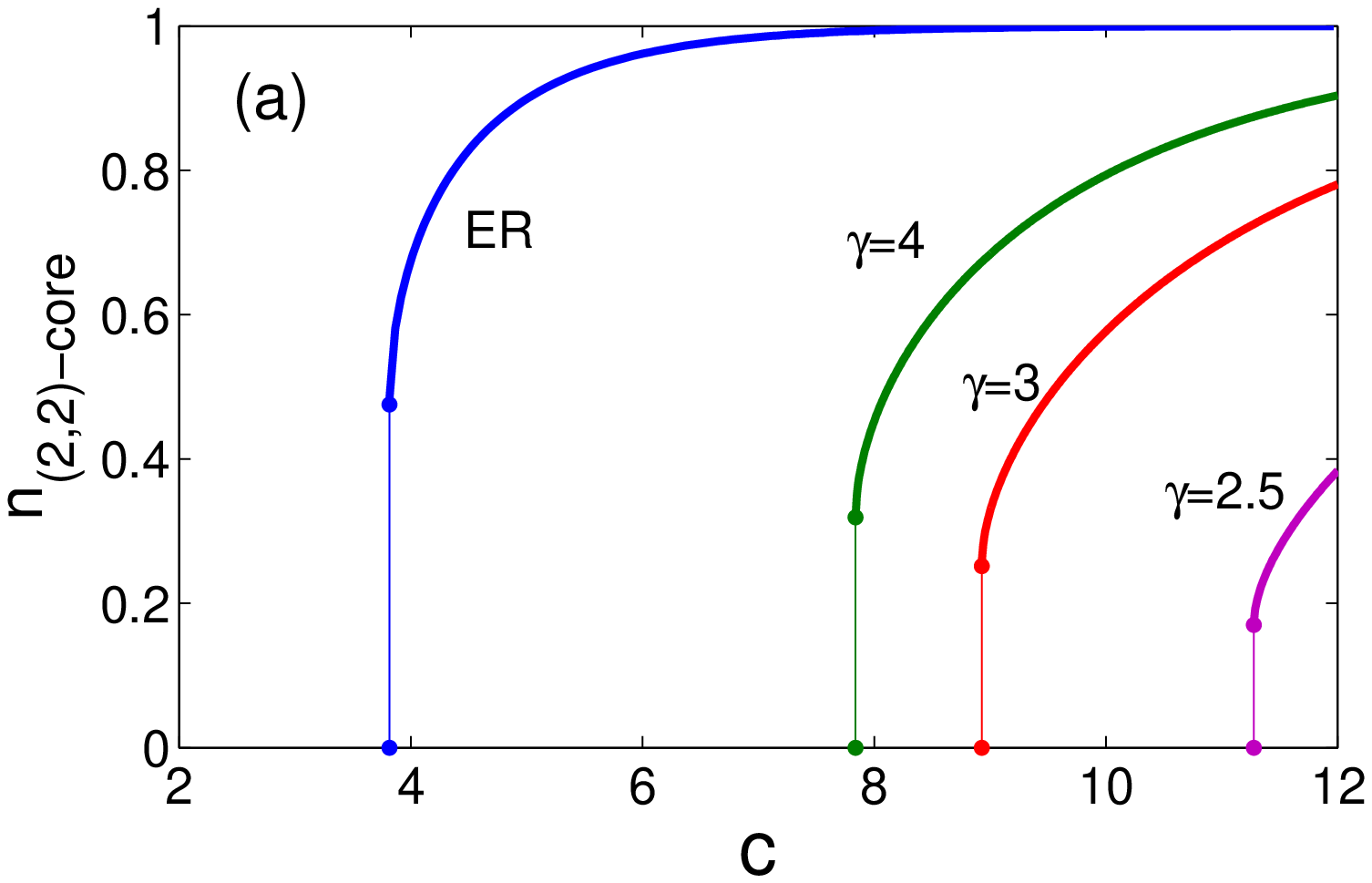}}
\scalebox{0.45}{\includegraphics[angle=0]{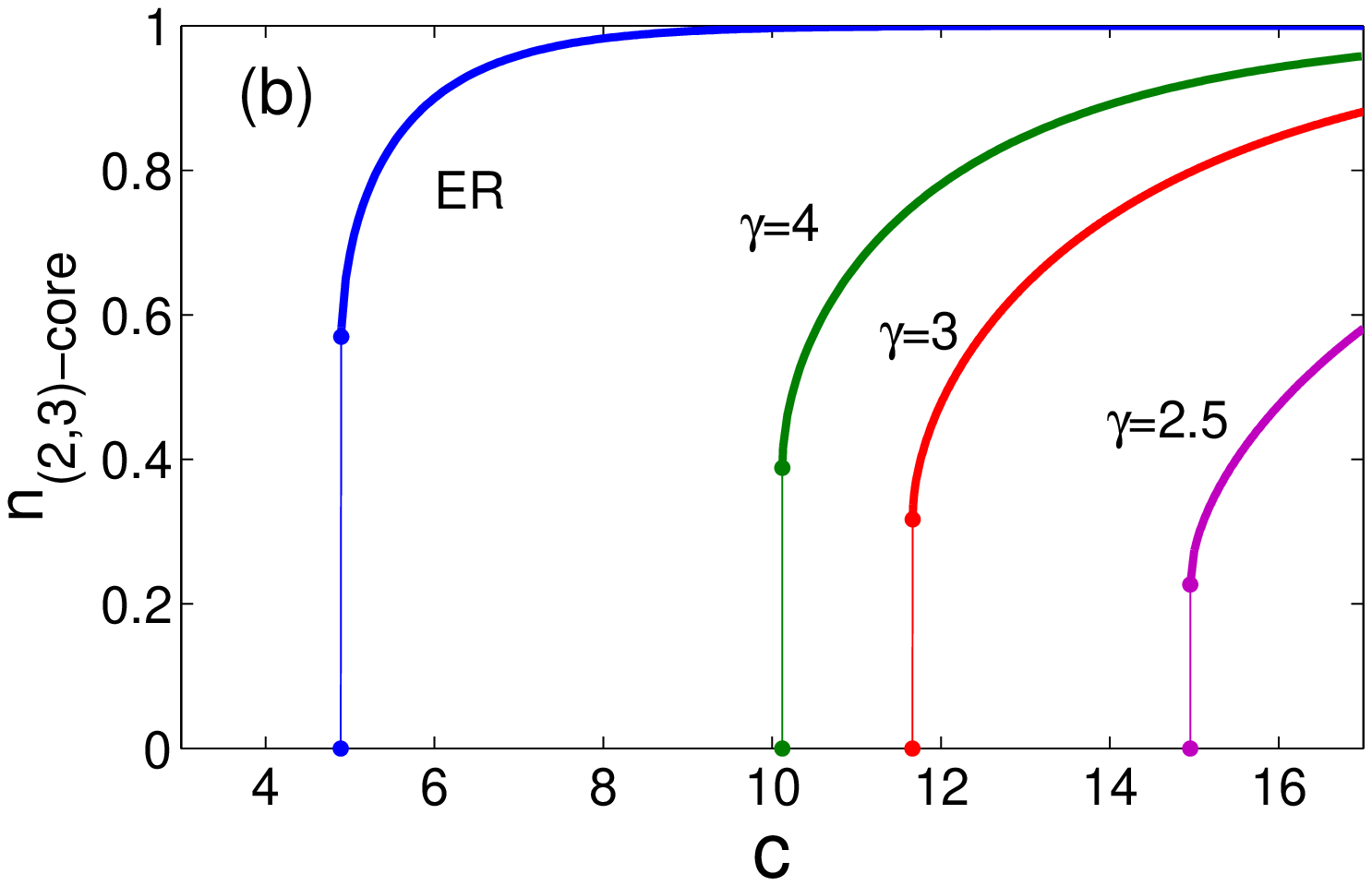}}
\end{center}
\caption{The relative size of $(a)$ $(2,2)-$core and $(b)$ $(2,3)-$core, for ER and asymptotically scale free networks, with different values of $\gamma$. For SF networks with smaller $\gamma$, the critical point shifted to higher mean degree values.
}
\label{f4}
\end{figure}
%%%%%%%%%%%%%%%%%%%%%%%%%%%%%%%%%%%%%%%%
%%%%%%%%%%%%%%%%%%%%%%%%%%%%%%%%%%%%%%%%

Next we consider asymptotically
%%SF
scale-free networks generated by the static  model with
\begin{equation}
P(q)=
\{
[\frac{\langle q\rangle(\gamma-2)}{2(\gamma-1)}]^{\gamma-1}
\Gamma(q-\gamma+1,
\frac{\langle q\rangle(\gamma-2)}{2(\gamma-1)}
)
\}
/\Gamma(q+1)\;,
\end{equation}
where $\Gamma(s)$ is the gamma function and $\Gamma(s,x)$ is the upper incomplete gamma function \cite{Goh:static}. This function in the large $q$ limit is asymptotically power law, $P(q) \sim q^{-\gamma}$ for $\gamma >2$. The generating function is $G(x)=(\gamma-1)E_n[(1-x)\frac{\langle q\rangle (\gamma-2)}{2(\gamma-1)}]$, where $E_n(x)=\int_{1}^{\infty}dy e^{-xy} y^{-n}$ is the exponential integral.

For different values of $\gamma$, Fig.~\ref{f4} shows the relative size of the $(2,2)$-core and the $(2,3)$-core and their corresponding  emergence points. The value of the jumps at the critical point, increases with increasing $\gamma$. Also, when the exponent $\gamma$ decreases, the transition point moves towards higher values of mean degree. In Fig.~\ref{f4} we compare the emergence of cores for scale-free and Erd\H{o}s--R\'enyi networks. As one can see,
%%the variation of
the dependency of the cores on $c$ for these networks is similar and, as expected, the curves with larger $\gamma$ approach to the result for Erd\H{o}s--R\'enyi networks.
The remarkable difference between Figs.~\ref{f3} and \ref{f4} for different scale-free multiplex networks points out that, similarly to ordinary $k$-cores, the $\textbf{k}$-cores essentially depend on the low-degree parts of the degree distributions.

%%%%%%%%%%%%%%%%
%%%%%%%%%%%%%%%
%%%%%%%%%%%%%

\section{Real multiplex networks}
\label{s3}

%%%%%%%%%%%%%%%%%%%%%%%%%%%%
%%%%%%%%%%%%%%%%%
\begin{figure}[t!]
\begin{center}
\scalebox{0.22}{\includegraphics[angle=0]{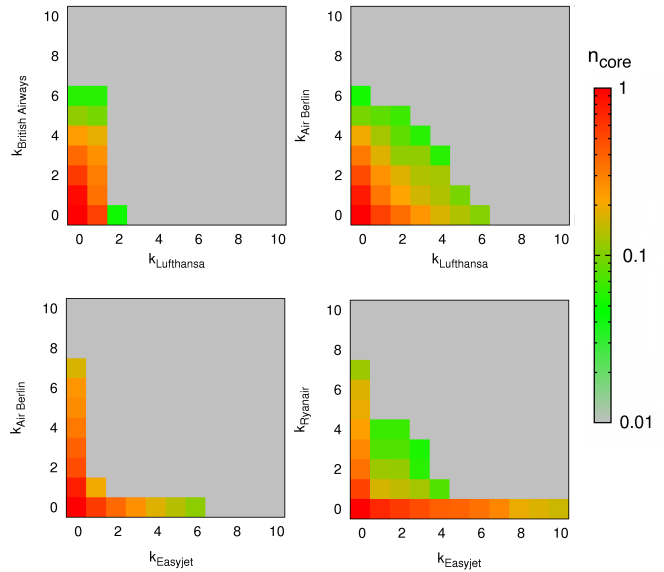}}
\end{center}
\caption{The relative size of $\textbf{k}$-cores of BritishAirway-Lufthansa, AirBerlin-Lufthansa, AirBerlin-EasyJet and Ryanair-EasyJet multiplex transportation networks, for different values of $k_a$ and $k_b$.
}
\label{f5}
\end{figure}
%%%%%%%%%%%%%%%%%%%%%%%%%%%

As an example of real multiplex networks, we consider air-transportation networks, in which vertices represent airports and edges direct flight connections between airports. Since flights can be operated by different airlines, a detailed representation of this kind of systems is given by multiplexes \cite{cardillo}. For the sake of simplicity we consider air-transportation multiplexes composed of two types of edges (two different airlines). Among the possible choices for composing these multiplexes, we have considered both low-cost airlines and major carriers. The main difference between the organization of these airlines is that low-cost companies diversify their main airports across and their goal is mainly driven by the economic growth. Instead major carriers are typically associated to one country as they are originally designed to serve the national and international mobility of the corresponding citizens. These major airlines are thus designed following the so-called {\em hub-and-spoke} structure in which one airport (the hub) is surrounded by many low degree vertices forming a kind of star-like graph.

We have  considered three different types of multiplexes comprising: {\em (i)} two low-cost airlines, {\em (ii)} two major airlines, and {\em (iii)} one low-cost and one major airline. In particular, we have considered these combinations: $EasyJet-Ryanair$ and $EasyJet-AirBerlin$ (combination of two low-cost airlines), $Lufthansa-British Airways$ (combination of two major airlines) and $Lufthansa-AirBerlin$ (combination of low-cost and major airlines).

Fig.~\ref{f5} shows the sizes of $\textbf{k}$-core of these networks for different values of $k_a$ and $k_b$. For the case $EasyJet-Ryanair$ and $Lufthansa-AirBerlin$ (which are two airlines
%%are
operating in the same country), we obtain more central cores, since the combined companies have many common vertices with a large number of connections for each type of edges. On the other hand, for $EasyJet-AirBerlin$, the common vertices have a few connections for each type of edges, which are removed in the first steps of $\textbf{k}-$core algorithm. Hence,
%%there are
this multiplex network has only a few cores. Similarly, the $Lufthansa-British Airways$ multiplex network has a few cores, since these two major airlines operate from different countries and thus the connectivity of the overlapping vertices is very different.
%%%%%%%%%%%%%%%%%%%%%%%%%%%%%%%%%
\begin{figure}[t!]
\begin{center}
\scalebox{0.38}{\includegraphics[angle=0]{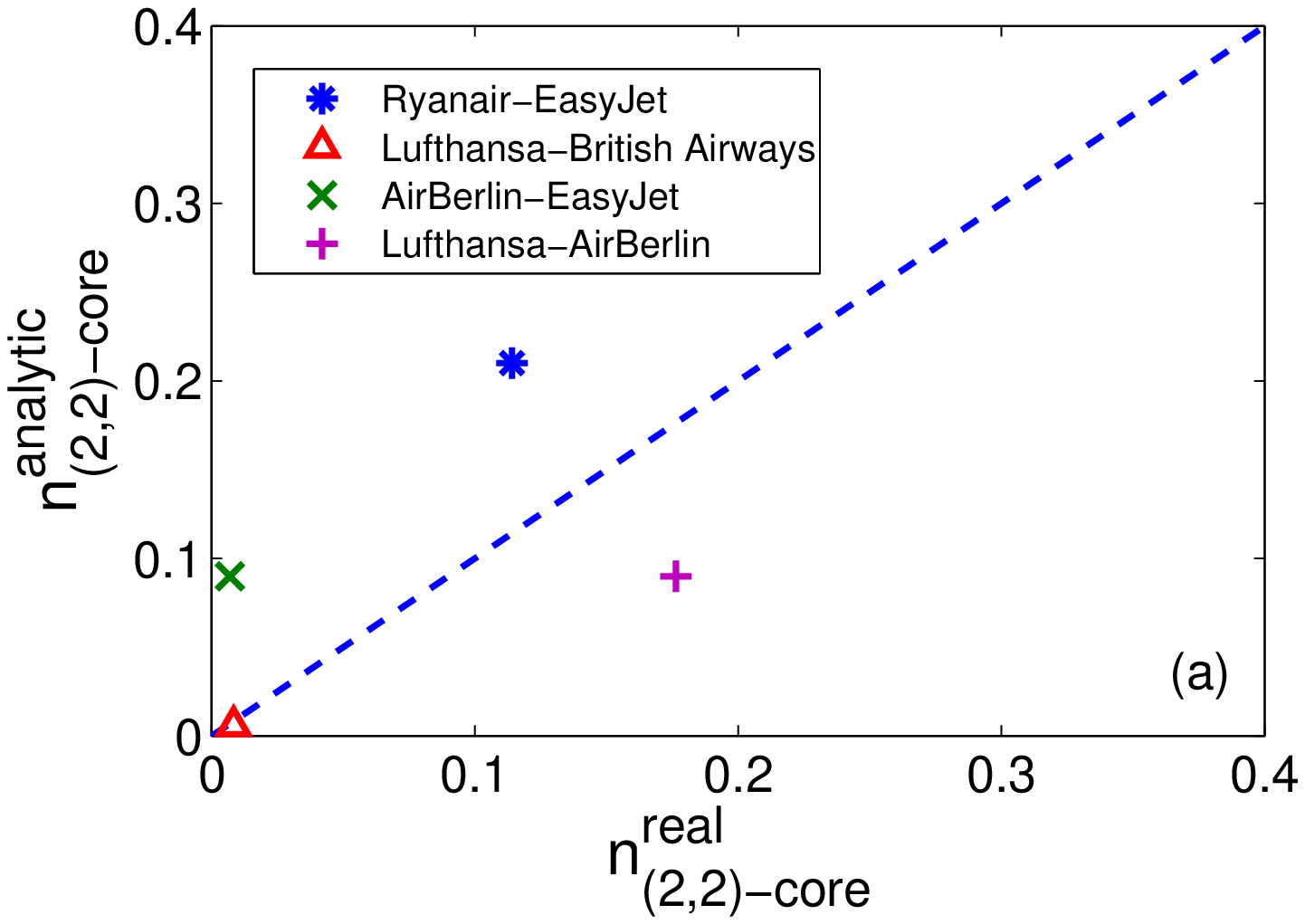}}
\scalebox{0.38}{\includegraphics[angle=0]{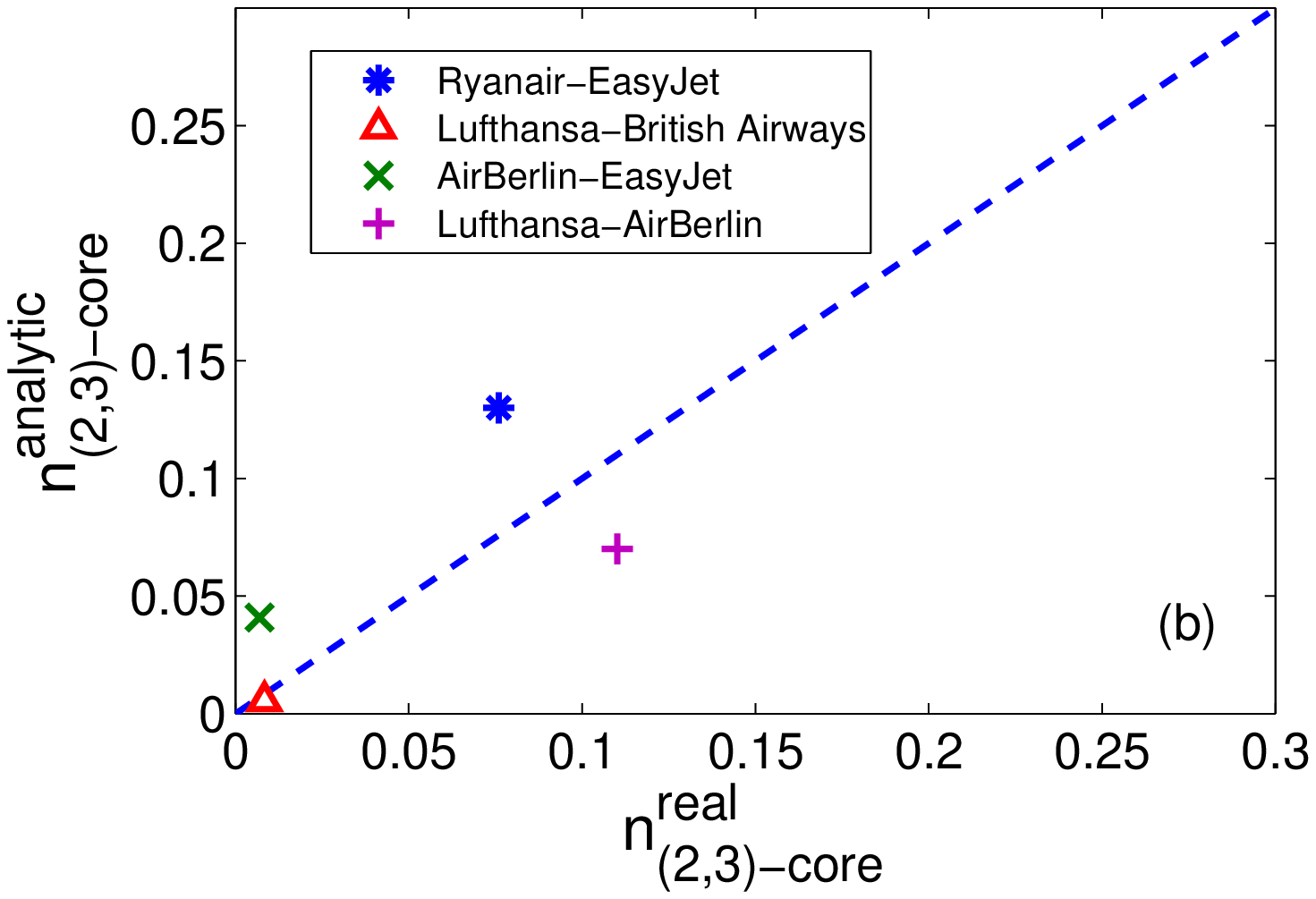}}
\end{center}
\caption{Comparison between $\textbf{k}$-core sizes obtained from empirical data and our theory for the relative sizes of
%%viable
$(a)$ the $(2,2)$-core and $(b)$ the $(2,3)$-core of transportation multiplex networks. Theoretical values are calculated for uncorrelated counterparts of the corresponding empirical multiplex networks.}
\label{f6}
\end{figure}

%%%%%%%%%%%%%%%%%%%%%%%%%%%%%%%%%
In Fig.~\ref{f6}, we have compared the size of the
%%viable
$(2,2)$-core and the $(2,3)$-core for air-transportation multiplex networks with the corresponding analytical results, obtained from Eqs.~(\ref{eq4})--(\ref{eq5}) in which we made use of the degree distributions of the empirical multiplex networks. As one can see, in some cases there is a noticeable difference between theory and reality, which arises from clustering, degree-degree correlations, and structural motifs in real-world
networks, which we did not take into account. Furthermore, the effects of overlapping between edges from different layers in the real-world multiplex networks can also be significant as noted in \cite{Cellai:clz13,Halu:hmb14}.

%%%%%%%%%%%%%%%%%%%%%%
%%%%%%%%%%%%%%%%%%%%%%

\section{Conclusion}
\label{s4}

%%In summary, w
In this work we have generalized the theory of $\textbf{k}$-core percolation to the multiplex networks. Our aim is to present a pruning algorithm for the $\textbf{k}$-core decomposition of multiplex networks that may be useful for describing the topological structure of these networks. To this aim, we introduced a $\textbf{k}$-viable component and showed that the
%%giant
$\textbf{k}$-core is the giant $\textbf{k}$-viable component if $k_i\geq 2$ for each type of edges. In this case, for each two vertices in the $\textbf{k}$-viable component there are at least $k_i$ paths, following only edges of type $i$, joining them.

We have analytically solved the $\textbf{k}$-core percolation problem for multiplex networks with arbitrary degree distributions. In the particular case of $(1,1)$-core, the transition is continuous. This is not in contradiction with previous results for ordinary percolation on multiplex networks,
%%in the sense that
because in this case the $(1,1)$-core is not a $\textbf{k}$-viable component. We showed however that the transitions for
%%the size of other
higher cores are hybrid.
Noticeably, although the $(1,2)$ and $(2,1)$-cores are not viable, they also display hybrid transitions.
%%We also observed difference between the paths connecting vertices in the $(1,2)$ and $2,1$-cores and those within the higher cores. The latter are viable.
%%become hybrid if $\sum_i k_i \geq 3$.
Moreover we found that the $(k_a,k_b,\ldots)$-core on uncorrelated multiplex networks has a higher threshold than the $k_i$-core on the counterpart single networks. Hence, we conclude that multiplex networks are less robust compared to their counterpart single networks,
%%against
if we analyse the destruction of the $\textbf{k}$-cores induced by random removal of vertices.

In summary, the $\textbf{k}$-core problem for the multiplex networks turns out to be essentially richer than the $k$-core problem for single networks. The pruning algorithm allows one to extract the $\textbf{k}$-cores in the multiplex networks in an easier way than their viable components. So the $\textbf{k}$-core decomposition of these complex networks is algorithmically efficient.
In the analytical framework presented here, we focused on uncorrelated and locally tree-like multiplex networks, ignoring the overlap of different types of edges, clustering, and correlations. Our empirical data analysis has revealed that these features may be significant for the sizes and organization of $\textbf{k}$-cores. We suggest that our theory could be extended to consider the case of complex multiplex networks with diverse structural correlations.
%%of the real data offers improvements for understanding of $\textbf{k}-$core percolation on real-world multiplex networks, which is a challenge for future work.
%%%%%%%%%%%%%%%%%%%%%%%%
%%%%%%%%%%%%%%%%%%%%%%%%%%%
%%%%%%%%%%%%%%%%%%%%%%
\begin{acknowledgments}
This work was partially supported by the Portuguese FCT under projects
%%projects PTDC:
%%FIS/108476/2008, SAU-NEU/103904/2008,
PTDC/MAT/114515/2009, and PEst-C/CTM/LA0025/2011; the Spanish MINECO under projects FIS2011-25167 and FIS2012-38266-C02-01; and the FET IP Project MULTIPLEX (317532). JGG is supported by the Spanish MINECO through the Ramon y Cajal program.
%%, and also by the SOCIALNETS EU project.
\end{acknowledgments}
%%%%%%%%%%%%%%%%%%%%%%%
%%%%%%%%%%%%%%%%%%%%%%%%
%%%%%%%%%%%%%%%%%%%%%%%%
%%%%%%%%%%%%%%%%%%%%%%%%%%

\end{document}